\documentclass[12pt,runningheads,a4paper]{amsart}
\usepackage{amssymb}
\usepackage{amsmath}

\begin{document}
\title{Spectral Function and Kinetic Equation for Normal Fermi
Liquid}
\author[M. Arshad${}^1$, A. S. Kondratyev${}^{1,2}$, I. Siddique${}^1$]
{M. Arshad${}^1$, A. S. Kondratyev${}^{1,2}$, I. Siddique${}^1$}
\address{${}^1$School of Mathematical Sciences, GC University\\
68-B New Muslim Town, Lahore, Pakistan. e-mail:
\ m.arshad.77@gmail.com\\
imransmsrazi@gmail.com
\\${}^2$Department of Physics, Herzen State Pedagogical
University\\Moika River Embankment, 48 191168 St. Petersburg,
Russia. e-mail: kondrat98926@yahoo.com}
\maketitle

\begin{abstract} On the basis of the Kadanoff-Baym (KB) version of the
time-dependent Green's function method, a new $Ansatz$ for the
approximation of a spectral function is offered. The $Ansatz$
possesses all the advantages of quasiparticle and extended
quasiparticle approximations and satisfies the KB equation for a
spectral function in the case of slightly nonequilibrium system when
disturbances in space and time are taken into consideration in the
gradient approximation. This feature opens opportunities for the
microscopic derivation of the Landau kinetic equation for the
quasiparticle distribution function of the normal Fermi liquid and
provides the widening of these equation's temperature range of
validity.\\
\\keywords: {Spectral function, normal Fermi liquid, quasiparticle
distribution, density matrix.}
\\PACS number(s): 21.65.+f, 72.10.Bg, 72.20Dp.
\end{abstract}

\section{Introduction}
 The only microscopic theory which is
capable of describing both the statistics and dynamics in a
comprehensive way is the Green's function approach developed in
different varieties for equilibrium and nonequilibrium problems and
for zero and finite temperatures. The initial emphasis in
application of Green's function techniques was to understand the
properties of normal condensed matter systems, superconductors, and
superfluids.

Microscopic models used for a description of quantum interacting
many-body systems involve spectral functions which play a central
role in the formalism. Spectral functions are fundamental in
describing the nuclear correlations, electrons correlations in
metals and semiconductors, and in many other equilibrium and
nonequilibrium properties. The most general approach to the problem
is based on the real-time Green's function formalism of Martin and
Schwinger, further developed by Kadanoff and Baym \cite{1}. Keldysh
\cite{2} developed a diagram version of the theory for
nonequilibrium systems equivalent to the  Kadanoff-Baym (KB)
approach. The KB equations were used in several contributions to
this evolving field with applications to nuclear matter \cite{3}, to
one- and two-band semiconductors \cite{4,5}, to electron plasmas
\cite{6}--\cite{8}, etc. The various approximations of the KB
equations differ essentially by the reduction schemes of the
two-time Green's functions to the reduced equilibrium density matrix
and to the quasiparticle distributions. There are several papers on
spectral functions published during the past years using different
methods and approximations \cite{9}--\cite{13}. The nonequilibrium
extension of the KB formalism has been recovered within the
quasiparticle approach to kinetic equation for weakly interacting
particles and referred to as a modified KB $Ansatz$
\cite{22}--\cite{24}.

In a variety of works beginning with \cite{1}, the spectral function
was approximated by a delta function of energy peaked at the
quasiparticle energy. Such a simple approximation falls short of
describing many important features of the systems under
consideration. Improved spectral functions were necessary and some
variants were offered in \cite{11,12,23}. The extended quasiparticle
(EQP) approximation was introduced and it was shown to remedy some
of the faults of the simpler quasiparticle (QP) approximation . Thus
the EQP approximation was a considerable improvement of the
formalism and it was used to compare the zero-temperature case of
the KB approach with Brueckner theory of nuclear matter
\cite{11,12}. The EQP approximation can be modified slightly while
maintaining its simplicity and renormalization property. One
possible modification of EQP was offered in \cite{12}, but it was
not even mentioned among possible forms of spectral functions in
\cite{13}.

Further development of the KB theory was performed in several
different directions. Numerical solutions of the KB equations were
presented in \cite{14}. The extended quasiparticle picture which was
first offered in \cite{15}, was developed for small scattering rates
in \cite{16}--\cite{18}. The separation of pole and off-pole parts
of a spectral function was discussed in detail in \cite{17}. The
nonlocal quasiparticle kinetic equation for the momentum-, space-,
and time-dependent distribution function, which has the form of a
Boltzmann equation with the quasiparticle energy, was derived in
\cite{18}. The generalization of the KB equations to the case of
arbitrary initial correlation was performed in \cite{19}.
Correlation effects related to the collision integral were shown to
cause a damping of the spectral function \cite{20}. However, for the
present analysis (see below), like in some other cases \cite{21}, a
detailed self-consistent microscopic treatment of the correlation
effects on the spectral function can be avoided.

Besides other applications, the spectral function was used for the
derivation of the kinetic equation of the phenomenological Landau
normal Fermi liquid theory. The initial derivation of this equation
was produced by Kadanoff and Baym on the basis of the quasiparticle
approximation for the spectral function and was continued by some
followers who used the extended quasiparticle approximation. In both
cases, the second Poisson bracket in the right side of the KB
generalized kinetic equation could not be eliminated in a lawful
mathematical way. This was the reason of the narrowing of the
temperature range of validity of Landau's equation.

This paper is devoted to the explanation of the successfulness of
the phenomenological Landau theory of normal Fermi liquid and its
applicability far beyond its temperature range of validity
established on the basis of quasiparticle and extended quasiparticle
$Ans\ddot{a}tze$ for the spectral function. We offer another
approximation for the spectral function which possesses the
advantage typical for QP and EQP approximations but has an
additional property to satisfy exactly the KB equation for the
spectral function in the case of slightly nonequilibrium systems. It
makes this form of the spectral function preferable in the
nonequilibrium systems and opens certain opportunities for the
widening of the range of validity of Landau Fermi-liquid kinetic
equation.

The following section contains a concise presentation of the
necessary formulas of the KB theory for both equilibrium and
slightly nonequilibrium systems which will be used in Sec. $3$ for a
comparison of three different approximations for the spectral
function. Section $4$ contains the analysis of the question that
what approximation for a spectral functions suits better to the
strict results of Kadanoff and Baym for the nonequilibrium case and
deals with the Landau kinetic equation for the quasiparticles in the
normal Fermi liquid which is proved to be valid in a wider
temperature region than it was considered in the initial derivation
on the basis of the KB formalism. In Sec. $5$ a discussion and
summary  are presented.

\section{Main formulas of the KB formalism}

We will introduce the main results of the KB theory for fermion
systems \cite{1} keeping in mind the questions which will be
discussed below. The KB formalism leads to the following general
expression for the one-particle spectral function $a(\vec{p}\omega)$
of a system in equilibrium:
\begin{equation}
a(\vec{p}\omega)=\frac{\Gamma(\vec{p}\omega)}{[\omega-E^{HF}(\vec{p})-
\text{Re}\ \sigma_c(\vec{p}\omega)]^2+
\frac{\Gamma^2(\vec{p}\omega)}{4}},
\end{equation}
where $E^{HF}(\vec{p})$ is a one-particle energy in the Hartree-Fock
approximation. Real and imaginary $(\Gamma)$ parts of the
correlation self-energy function $\sigma_c$ are related to each
other through the Hilbert transform,
\begin{equation}
\text{Re}\
\sigma_c(\vec{p}\omega)=P\int_{-\infty}^{\infty}\frac{d\omega^{'}}{2\pi}
\frac{\Gamma(\vec{p}\omega^{'})}{\omega-\omega^{'}} .
\end{equation}
Here, $P$ refers to a principal value integration.

The spectral function satisfies the exact sum rule:
\begin{equation}
\int_{-\infty}^{\infty}\frac{d\omega}{2\pi}a(\vec{p}\omega)=1,
\end{equation}
for all the values of $\vec{p}$. This result follows directly from
the commutator relations for field operators and can serve as a
keystone for the checking of the validity of all approximations for
the spectral function (1).

In the Hartree-Fock approximation, when the correlation self-energy
function $\text{Re}\ \sigma_c(\vec{p}\omega)=0$, the spectral
function turns to be a delta function of the Hartree-Fock energy:
\begin{equation}
a_{HF}(\vec{p}\omega)=2\pi\delta[\omega-E^{HF}(\vec{p})],
\end{equation}
and the sum rule (3) is trivially satisfied identically for any
value of $\sigma^{HF}(\vec{p})$. The simplest approximation in the
case when $\text{Re}\ \sigma_c(\vec{p}\omega)\neq0$ is the so called
(QP) approximation:
\begin{equation}
a_{QP}=2\pi Z(\vec{p})\delta(\omega-E(\vec{p})) ,
\end{equation}
where $E(\vec{p})$ is the solution of the equation:
\begin{equation}
E(\vec{p})=E^{HF}(\vec{p})+\text{Re}\ \sigma_c[\vec{p},E(\vec{p})],
\end{equation}
and the renormalizing factor $Z(\vec{p})$ is given by the expression
\begin{equation}
Z^{-1}(\vec{p})=1-\frac{\partial \text{Re}\
\sigma_c(\vec{p}\omega)}{\partial{\omega}}\Big|_{\omega=E(\vec{p})}.
\end{equation}
In the QP approximation, the sum rule (3) reads
\begin{equation}
\int_{-\infty}^{\infty}\frac{d\omega}{2\pi}a_{QP}(\vec{p}\omega)=Z(\vec{p}).
\end{equation}
A severe drawback with $a_{QP}$ is that it normalizes to
$Z(\vec{p})$ rather than to 1 as in Eq. (3). We will see below how
this deficiency is removed by the extended quasiparticle
approximation \cite{11}--\cite{13} and by other more advanced
approximations.

In the case of slowly varying in space and time disturbances, after
the transition to Wigner coordinates,
\begin{equation}
\vec{R}=\frac{1}{2}(\vec{r_1}+\vec{r_{1^{'}}}),\ \hspace{.1cm}
\vec{r}=\vec{r_1}-\vec{r_{1^{'}}} \hspace{.2cm},\ \hspace{.4cm}
T=\frac{1}{2}(t_1+t_{1^{'}}),\ \hspace{.1cm}t=t_1-t_{1^{'}},
\end{equation}
and the performance of the Fourier transform with respect to
$\vec{r},t,$ all the quantities entering the theory are considered
to be the functions of $\vec{p},\omega,\vec{R},T,$ for example,
$$a=a(\vec{p}\omega;\vec{R}T).$$
If we take into account only the first derivatives with respect to
slowly varying quantities $\vec{R}$ and $T$ in the KB equations for
the correlation functions, we come to the following equation for the
spectral function $a(\vec{p}\omega;\vec{R}T)$ \cite{1}:
\begin{equation*}
[\omega-E^{HF}(\vec{p};\vec{R}T)-\text{Re}\
\sigma_c(\vec{p}\omega;\vec{R}T),a(\vec{p}\omega;\vec{R}T)]
\end{equation*}
\begin{equation}
+[\text{Re
}g(\vec{p}\omega;\vec{R}T),\Gamma(\vec{p}\omega;\vec{R}T)]=0,
\end{equation}
and to the generalized KB kinetic equation for the correlation
function $g^<(\vec{p}\omega;\vec{R}T)$:
\begin{equation*}
[\omega-E^{HF}(\vec{p};\vec{R}T)-\text{Re}\
\sigma_c(\vec{p}\omega;\vec{R}T),g^<(\vec{p}\omega;\vec{R}T)]
\end{equation*}
\begin{equation}
+[\text{Re }\
g(\vec{p}\omega;\vec{R}T),\sigma^<(\vec{p}\omega;\vec{R}T)]=
(\sigma^<g^>-\sigma^>g^<)(\vec{p}\omega;\vec{R}T).
\end{equation}
Here, $[A,B]$ is the so called generalized Poisson bracket, defined
by the expression:
\begin{equation}
[A,B]=\frac{\partial{A}}{\partial{\omega}}\frac{\partial{B}}{\partial{T}}-
\frac{\partial{A}}{\partial{T}}\frac{\partial{B}}{\partial{}\omega}-
\nabla_{\vec{p}}A\ .\ \nabla_{\vec{R}}B+\nabla_{\vec{R}}A\ .\
\nabla_{\vec{p}}B,
\end{equation}
$E^{HF}$ and $\text{Re}\ \sigma_c$ include the interaction with the
external field $U(\vec{R}T)$. The exact solution of Eq. (10) is
given by the expression
\begin{equation}
g(\vec{p}z;\vec{R}T)=[z-E^{HF}(\vec{p};\vec{R}T)-\text{Re}\
\sigma_c(\vec{p}z;\vec{R}T)]^{-1},
\end{equation}
where all the functions, such as $a(\vec{p}\omega;\vec{R}T)$,
$\text{Re}\ g(\vec{p}\omega;\vec{R}T)$, etc., are determined by the
same formulas as in the equilibrium case with all entering
quantities depending on $\vec{p},\omega,\vec{R},T$. For example,
\begin{equation*}
\text{Re}\
\sigma_c(\vec{p}\omega;\vec{R}T)=P\int_{-\infty}^{\infty}\frac{d\omega^{'}}{2\pi}
\frac{\Gamma(\vec{p}\omega^{'}; \vec{R}T)}{\omega-\omega^{'}}.
\end{equation*}
In fact, the solution (13) gives almost the same evaluation of the
spectral function $a$ as in the equilibrium case:
\begin{equation*}
a(\vec{p}\omega;\vec{R}T)=\frac{\Gamma(\vec{p}\omega;\vec{R}T)}{[\omega-E^{HF}
(\vec{p};\vec{R}T)-\text{Re }\
\sigma_c(\vec{p}\omega;\vec{R}T)]^2+\frac{\Gamma^2(\vec{p}\omega;\vec{R}T)}{4}}.
\end{equation*}

 Since the solution (13) is of exactly the same form
as the equilibrium solution, it must reduce to the equilibrium
solution as $T\longrightarrow{-\infty}$. Thus, it satisfies the
initial condition on the equation of motion. This result means that
the approximation for the nonequilibrium spectral function can be
written in the same form as in the equilibrium case.

  Equation (11) provides an exact description of the response to
slowly varying disturbance. All the quantities appearing in this
equation may be expressed in terms of the correlation functions
$g^<$ and $g^>$. In particular, $\sigma^<$ and $\sigma^>$ are
defined by a Green's function approximation that gives the
self-energy in terms of $g^<$ and $g^>$.

\section{Approximations for the spectral function}

We will consider three different approximations for the spectral
function which we will call $Ansatz$ 1, $Ansatz$ 2, and $Ansatz$ 3
correspondingly. Thus, we have
\begin{eqnarray}
\text{$Ansatz$\ 1}:\quad&  a_{EQP} = 2\pi
Z(\vec{p})\delta(\omega-E)+P\frac{\Gamma(\omega)}{(\omega-E)^2}; \\
\text{$Ansatz$ 2}:\quad &  a_{iQP}=2\pi Z(\vec{p})\delta(\omega-E)+
Z(\vec{p})P\frac{\Gamma(\omega)}{(\omega-E)^2};\\
\text{$Ansatz$ 3}:\quad & a_{i}=2\pi
Z(\vec{p})\delta(\omega-E)+Z^2(\vec{p})
P\frac{\Gamma(\omega)}{(\omega-E)^2}.
\end{eqnarray}
Here, $E$ is determined by Eq. (6).

The $Ansatz$ 1 corresponds to the extended quasiparticle
approximation which was introduced, discussed, and used for
numerical calculation in \cite{11}--\cite{13}. The first  form of
this $Ansatz$ was offered in \cite{25}. The $Ansatz$ 2 as it was
already mentioned above, offered in \cite{12}, but not discussed and
was never used for the numerical calculation and was not even
mentioned in \cite{13}. A little bit different in technical details
but the same in principle way of introducing the $Asatz$ 1 was
offered in \cite{23}. We would like to emphasize that the forms (14)
and (15) of the spectral function were constructed in
\cite{11}--\cite{13}, \cite{23} on the basis of the general
expression (1) and were obtained by means of the Taylor expansion in
powers of $\Gamma$ in the frame of different approximations. In
reality, there does not exist a mathematically strict correct form
for the expansion of (1) in power series of $\Gamma({\omega})$ which
starts with the delta function when $\Gamma({\omega})=0$. Thus,
expression (14) and (15) should be considered as some true-like
approximate forms, as it is recognized in \cite{23}, which have
better qualities compared with the quasiparticle $Ansatz$ (5). In
particular, Eqs. (14) and (15) obey the sum rule (3) (see below).
Neither can be obtained in a strict way the improved $Ansatz$ 3 $
a_i$ offered by us. The origin of the possibility of such $Ansatz$
on equal terms with expressions (14) and (15) can be shown on the
basis of the following consideration.

We start with a well known relation of the Fourier transform in the
case of a constant value of $c$:
\begin{equation}
\int_{-\infty}^{\infty}e^{-|t|c}e^{\imath{tx}}dt=\frac{2c}{c^2+x^2}\hspace{.1cm}
 ,\hspace{.5cm} c>0.
\end{equation}
Expanding the first exponent in the left side of Eq. (17) in Taylor
series, we get
\begin{equation}
\frac{2c}{c^2+x^2}=\int_{-\infty}^{\infty}\Big(1-c|t|+\frac{{c^2t^2}}{2!}-...\Big)
e^{\imath{tx}}dt.
\end{equation}
Now we use the formulas equivalent to those represented in
\cite{26}--\cite{28}:
\begin{eqnarray}
\int_{-\infty}^{\infty}t^{(2n)}e^{\imath{tx}}dt&=&
2\pi (-\imath)^{2n}\delta^{(2n)}(x) ,\hspace{.5cm} n=0,1,2,... \\
\int_{-\infty}^{\infty}|t|^{(2n+1)}e^{\imath{tx}}dt &=&
-2\sin\Big((2n+1)\frac{\pi}{2}\Big)(2n+1)!\frac{1}{|x|^{2n+2}},\\
\nonumber\hspace{.1cm}n=0,1,2,...\ .
\end{eqnarray}
If the quantity $\Gamma{(\omega)}$ in Eq. (1) is constant, then the
expression (18) would lead to a strict correct expansion of spectral
function $a(\omega)$ in terms of the power series of $\Gamma$.
However, $\Gamma{(\omega)}$ cannot be constant due to the dispersion
relation (2). In the case of $\Gamma{(\omega)}$ depending on
$\omega$, one can rely only on the first two terms of the expansion:
the delta function independent of $\Gamma{(\omega)}$ and the term
proportional to $\Gamma{(\omega)}$. Then one should notice that the
quantity $x$ in Eq. (17) or (18) corresponds to
$\omega-E^{HF}(\vec{p})-\text{Re}\ \sigma_c(\vec{p}\omega)$ in the
formula (1). Taking into account the relation (6), it becomes clear
with the precision to the second derivative $\frac{\partial^2\
\text{Re}\ {\sigma_c}}{\partial{\omega^2}}$, the term of the
expansion (18) proportional to  $\Gamma{(\omega)}$ involves $Z^2(p)$
in the numerator, and we come to the formula (16). Indeed, in the
mentioned approximation we have
\begin{equation*}
\omega-E^{HF}(\vec{p})-\text{Re }\
\sigma_c(\vec{p}\omega)=Z^{-1}(\vec{p})[\omega-E(\vec{p})].
\end{equation*}

 Now we should compare the expressions (14)-(16) on
the basis of strict results obtained in the KB theory. We will check
the correspondence of these expressions to the sum rule (3),  the
second sum rule on the basis of energy considerations and to Eq.
(10) which determines the spectral function in the case of slowly
varying in space and time disturbances. We start with the sum rule
(3). Substitute $Ansatz$ 1 given by Eq. (14) to Eq. (3) and we get
\begin{equation}
\int_{-\infty}^{\infty}\frac{d\omega}{2\pi}a_{EQP}=Z(\vec{p})+
P\int_{-\infty}^{\infty}\frac{d\omega}{2\pi}\frac{\Gamma(\omega)}{(\omega-E)^2}.
\end{equation}
Due to the dispersion relation (2) we have
\begin{equation}
\frac{\partial\ \text{Re}\
\sigma_c(\omega)}{\partial{\omega}}\mid_{E}=
-P\int_{-\infty}^{\infty}\frac{d\omega}{2\pi}\frac{\Gamma(\omega)}{(\omega-E)^2},
\end{equation}\\and Eq. (21) becomes
\begin{equation}
\int_{-\infty}^{\infty}\frac{d\omega}{2\pi}a_{EQP}=\frac{1}{1-\frac{\partial\
\text{Re}\
\sigma_c(\omega)}{\partial{\omega}}\mid_{E}}-\frac{\partial\
\text{Re}\ \sigma_c(\omega)}{\partial{\omega}}\mid_{E} \approx 1,
\end{equation}\\with the precision to
$(\frac{\partial\ \text{Re }\
{\sigma_c}}{\partial{\omega}})^2$.\\The $Ansatz$ 2 gives
\begin{equation}
\int_{-\infty}^{\infty}\frac{d\omega}{2\pi}a_{iQP}=
Z(\vec{p})\Big(1+P\int_{-\infty}^{\infty}\frac{d\omega}{2\pi}
\frac{\Gamma(\omega)}{(\omega-E)^2}\Big).
\end{equation}
Taking into account the relation (22), we see that in the case under
consideration the integral (24) is equal to 1.

Substituting Eq. (16) into Eq. (3) and taking into account (22), one
gets
\begin{equation}
\int{\frac{d\omega}{2\pi}a_{i}}=Z(\vec{p})\Big(1-Z\frac{\partial\
\text{Re}\ \sigma_c(\omega)}{\partial{\omega}}\mid_{E}\Big)\approx
1,
\end{equation}
with the precision to $(\frac{\partial\ \text{Re}\
{\sigma_c{(\omega)}}}{\partial{\omega}})^2$.

Thus, formally the $Ansatz$ 2 is the best in the sense of sum rule
(3), but errors brought by $Ansatz$ 1 and $Ansatz$ 3 can be
neglected in the approximation under consideration.

We can check the expressions (14)-(16) also with the help of the
second sum rule \cite{13}:
\begin{equation}
\int_{-\infty}^{\infty}\frac{d\omega}{2\pi}\omega
a{(\vec{p}\omega)}=E^{HF}(\vec{p}).
\end{equation}

The first terms in the right side of Eqs. (14)-(16) give the same
result
\begin{equation}
\int_{-\infty}^{\infty}\frac{d\omega}{2\pi}\omega2\pi
Z(\vec{p})\delta(\omega-E)=Z[E^{HF}+\text{Re}\ \sigma_c(E)].
\end{equation}
The contributions of the second terms in right sides of Eqs.
(14)-(16) differ only in factors depending on the power of the
renormalizing factor Z. The corresponding integral in each case is:
\begin{eqnarray}
\nonumber
P\int_{-\infty}^{\infty}\frac{d\omega}{2\pi}\frac{{\omega}\Gamma(\omega)}{(\omega-E)^2}&=
&
P\int_{-\infty}^{\infty}\frac{d\omega}{2\pi}\frac{\Gamma(\omega)}{(\omega-E)}+
E P\int_{-\infty}^{\infty}\frac{d\omega}{2\pi}\frac{\Gamma(\omega)}{(\omega-E)^2} \\
&=& Z^{-1}[(1-Z)E^{HF}-(2Z-1)\text{Re}\ \sigma_c(E)].
\end{eqnarray}

The last equality in Eq. (28) follows with the help of the
expressions (2), (7), and (22). Now taking into account the
renormalizing factors standing with the second terms in Eqs.
(14)-(16), we get with the help of Eqs. (27) and (28) that the
second sum rule (26) is valid for all the $ans\ddot{a}tze$ (14)-(16)
with the same precision up to the terms of the order
$(\frac{\partial\ \text{Re}\
{\sigma_c{(\omega)}}}{\partial{\omega}})^2$ and in this sense they
are equivalent in principle. However, they turn to be not equivalent
in the sense of satisfying the Eqs. (10) and (11), although it is
mentioned in \cite{23} that the corresponding off-pole part of $g^<$
for the $Ansatz$ 1 (14) compensates a\ "dominant part of the
puzzling term $[\text{Re}\ g,\sigma^<]$" in the generalized kinetic
equation (11). However, represented in \cite{23} analysis did not
lead to mathematically lawful complete elimination of this puzzling
term. In the next section, we show that the $Ansatz$ 3 (16) solves
the problem and consequently is the best in the nonequilibrium case.

\section{Kinetic equation of the Landau theory
of the Normal Fermi Liquid}

According to the strict result (13), every approximate $Ansatz$ for
the spectral function should keep valid in the case of slowly
varying in space and time disturbances. The only difference occurs
due to the dependence of all quantities on $\vec{R}$ and T. When we
substitute Eqs. (14)-(16) to Eq. (10), the first terms in the right
side of these expressions give
\begin{equation*}
[\omega-E^{HF}(\vec{p})-\text{Re }\
\sigma_c(\vec{p}\omega;\vec{R}T),2\pi Z(\vec{p})\delta(\omega-E)]
\end{equation*}
\begin{equation}
=2\pi Z(\vec{p})\Big[(\omega-E)\Big(1-\frac{\partial\ \text{Re}\
{\sigma_c}}{\partial{\omega}}\mid_{E}\Big),\delta(\omega-E)\Big]=0,
\end{equation}
due to the property of the generalized Poisson bracket
\begin{equation}
[A,f(A)]=0.
\end{equation}
The second term in the right side of Eq. (14) leads to the
expression
\begin{equation*}
\Big[\omega-E^{HF}(\vec{p})-\text{Re}\
\sigma_c(\vec{p}\omega;\vec{R}T),\frac{\Gamma(\vec{p}\omega;\vec{R}T)}{(\omega-E)^2}\Big]
\end{equation*}
\begin{equation}
=Z^{-1}(\vec{p})\frac{1}{(\omega-E)^2}[\omega-E,\Gamma],
\end{equation}
with the precision to $\frac{\partial^2\ \text{Re }\
{\sigma_c{(\omega)}}}{\partial{\omega}^2}$. The second Poisson
bracket in Eq. (10) due to Eq. (13) gives with the same precision
the expression
\begin{eqnarray}
\nonumber[\text{Re }\
g(\vec{p}\omega;\vec{R}T),\Gamma{(\vec{p}\omega;\vec{R}T)}]=
Z(\vec{p})\Big[\frac{1}{(\omega-E)},\Gamma\Big]&\\
=-Z(\vec{p})\frac{1}{(\omega-E)^2}[\omega-E,\Gamma].&
\end{eqnarray}
Finally, collecting all the terms, we get that the left side of Eq.
(10) in the case under consideration is different for Eqs. (14)-(16)
and equals to

\begin{eqnarray}
\text{$Ansatz$ 1} &:& \Big[\frac{1}{(\omega-E)},\Gamma\Big][Z(\vec{p})-Z^{-1}(\vec{p})];\\
\text{$Ansatz$ 2} &:& \Big[\frac{1}{(\omega-E)},\Gamma\Big][Z(\vec{p})-1];\\
\nonumber \text{$Ansatz$ 3} &:&
\Big[\frac{1}{(\omega-E)},\Gamma\Big][Z(\vec{p})-Z(\vec{p})]=0.
\end{eqnarray}
The $Ansatz$ 3 satisfies Eq. (10) exactly. It means that this
$Ansatz$ is preferable for considering slightly nonequilibrium
systems when disturbances slowly vary in space and time. In
particular, it turns out that the $Ansatz$ 3 (16) opens
opportunities for the derivation of the Landau Fermi liquid's theory
kinetic equation for the quasiparticle distribution function. We
will see that the usual considerations about the smallness of the
quantity $\Gamma(\omega)$ are not necessary anymore and the puzzling
term $[\text{Re}\ g,\sigma^<]$ in Eq. (11) is canceled completely by
the off-pole part of the approximation (16).

From the pioneer work of Kadanoff and Baym, the derivation of the
kinetic equation of normal Fermi liquid theory was based on the
assumption of the smallness of the functions $\sigma^<$ and
$\sigma^>$ near the Fermi level $\mu$ of the system \cite{1,29}.
Indeed, defining a local occupation number
$f(\vec{p}\omega;\vec{R}T)$ by writing
\begin{equation}
g^<(\vec{p}\omega;\vec{R}T)=a(\vec{p}\omega;\vec{R}T)f(\vec{p}\omega;\vec{R}T),
\end{equation}
where in equilibrium, at zero temperature
\begin{equation}
f(\vec{p}\omega;\vec{R}T)\rightarrow f(\omega)=\begin{array}{ccc}
\end{array}
\left\{
\begin{array}{ll}
0 & \hbox{ \ \ for  \ \ $\omega>\mu$} \\
1 & \hbox{ \ \ for \ \ $\omega<\mu$}\ ,
\end{array}
\right.
\end{equation}
one comes to the relations
\begin{eqnarray}
\nonumber\sigma^>(\vec{p}\omega)=0 \ \ \ \ \hbox{for}
\ \ \ \omega<\mu&, \\
\sigma^<(\vec{p}\omega)=0 \ \ \ \ \hbox{for} \ \ \ \omega>\mu&.
\end{eqnarray}
The proof of these relations depends only the fact that\ \ $f=1$ \ \
\ for \ \ $\omega<\mu$ \ \ and  \ \  $f=0$ \ \  for \ \
$\omega>\mu$. Since we are assuming that $f$ has a similar behavior
in the slightly nonequilibrium case, it follows that
\begin{eqnarray}
\nonumber\sigma^>(\vec{p}\omega;\vec{R}T)=0 \ \ \ \ \hbox{for} \ \ \
\omega<\mu(\vec{R}T)&, \\
\sigma^<(\vec{p}\omega;\vec{R}T)=0 \ \ \ \ \hbox{for} \ \ \
\omega>\mu(\vec{R}T)&,
\end{eqnarray}
where $\mu(\vec{R}T)$ is a local chemical potential of the system.
For normal Fermi systems, $\sigma^>$ and $\sigma^<$ are continuous
functions at $\omega=\mu$. It means that $\Gamma=\sigma^>+\sigma^<$
is small near $\mu$. As it was shown in \cite{1}, these assumptions
about $f$ for $\omega$ appreciably greater or less than
$\mu(\vec{R}T)$ lead to a consistent solution of the generalized
kinetic equation (11). In such a situation when $\sigma^>$ and
$\sigma^<$ are both negligible for $\omega$ near $\mu$, Eq.  (11) in
collisionless approximation can be written in the form
\begin{equation}
[\omega-E^{HF}(\vec{p};\vec{R}T)-\text{Re}\
\sigma_c(\vec{p};\vec{R}T),a(\vec{p}\omega;\vec{R}T)f(\vec{p}\omega;\vec{R}T)]=0.
\end{equation}
The second Poisson bracket in the right side of Eq. (11) is dropped
out due to the discussed considerations about the smallness of
$\Gamma$ near $\mu$.

Using the quasiparticle $Ansatz$ (5) for the spectral function $a$,
one comes to the kinetic equation for the quasiparticle distribution
function $n(\vec{p},\vec{R}T)$ of the normal Fermi liquid theory
\cite{1}:
\begin{equation}
\frac{\partial n}{\partial T}+\nabla_{\vec{P}}E\ .\
\nabla_{\vec{R}}n-\nabla_{\vec{R}}E\ .\ \nabla_{\vec{p}}n=0,
\end{equation}
$$n(\vec{p};\vec{R}T)=f(\vec{p}\omega;\vec{R}T)\mid_{\omega=E(\vec{p};\vec{R}T)}.$$
The smallness of the function $\Gamma$ is a keystone in the above
considerations and such a smallness looks like a necessary condition
of the validity of Eq. (40), which is supposed to be valid only in
the region $\omega\approx\mu(\vec{R}T)$. However, it can be shown
that neglecting collisions, Eq. (40) stays valid with the precision
to $\Gamma^2$ if one uses $Ansatz$ 3 (16) for the spectral function
$a$ instead of the quasiparticle $Ansatz$ (5). In this case, the
collisionless equation (11) is written in the form
\begin{equation}
[\omega-E^{HF}-\text{Re}\ \sigma_c,af]+[\text{Re}\ g, \sigma^<]=0.
\end{equation}
The first term in all the $Ans\ddot{a}tze$ (14)-(16) which coincides
with the quasiparticle $Ansatz$ (5) leads directly to Eq. (40) as it
is shown in \cite{1}. The second Poisson bracket in Eq. (41) in the
linear approximation in $\Gamma$ gives the result
\begin{eqnarray}
\nonumber[\text{Re}\ g,\sigma^<]=\Big[\frac{1}{\omega-E^{HF}-
\text{Re}\ \sigma_c},\Gamma f\Big]=
Z\Big[\frac{1}{\omega-E},\Gamma f\Big]& \\
=Z[\Gamma f,\omega-E]\frac{1}{(\omega-E)^2}.&
\end{eqnarray}
It is easy to see that this term will be compensated by the second
term in the $Ansatz$ 3. Indeed, substituting this second term to the
first Poisson bracket in Eq. (41), one gets
\begin{equation}
Z^2\Big[\omega-E^{HF}-\text{Re}\ \sigma_c,\frac{\Gamma
f}{(\omega-E)^2}\Big]=Z [\omega-E,\Gamma f]\frac{1}{(\omega-E)^2},
\end{equation}
and this term compensates Eq. (42) in Eq. (41). Observe that neither
$Ansatz$ 1 nor $Ansatz$ 2 will lead to this result.

We would like to stress that all considerations based on the
relations (36)--(38) were necessary only for the sake of the
elimination of the second Poisson bracket in Eq. (41). The usage of
the $Ansatz$ 3 (16) makes the smallness of the function $\Gamma$ not
necessary. The quantity $\Gamma(\omega)$ can be finite. For
qualitative estimation of the precision, we can use the third term
in the expansion (18) which corresponds to $\Gamma^2(\omega)$.
Substituting the term proportional to $\Gamma^2(\omega)$ into Eqs.
(10) and (11), it is not difficult to show that in the gradient
approximation these equations are valid up to the terms of the order
$\Gamma^2(\omega)$.

This result means that the collective excitation spectrum of the
system which is determined by the kinetic equation (40) is not
changed when we take into considerations the term in the spectral
function proportional to $\Gamma$. At the same time, physical
quantities which are determined by the correlation function $g^<$
will obtain additional terms due to the second term in the
expression for the spectral function. In particular, the reduced
density matrix,
\begin{equation}
\rho(\vec{p};\vec{R}T)=\int_{-\infty}^{\infty}\frac{d\omega}{2\pi}g^<(\vec{p}\omega;\vec{R}T),
\end{equation}
will possess power tails while the quasiparticle distribution $n$
decrease exponentially with energy.

\section{Discussion and Summary}

The differences in the form of the expressions (14)--(16) come
through the ways they were obtained. When the $Ansatz$ for the
spectral function is obtained by a Taylor expansion of Eq. (1)
around the quasiparticle peak, it involves the necessity of the
conditions $\Gamma<<\text{Re}\ \sigma_c$ and $\frac{\partial\
{\text{Re}\ \sigma_c}}{\partial{\omega}}<<1$ \cite{11}-\cite{13}.
The approach (18) based on the Fourier transform demands another
condition $(\omega-E)\frac{\partial\ \ln\
\Gamma(\omega)}{\partial\omega}<<1$, which means that a relation
between $\Gamma$ and $\text{Re}\ \sigma_c$ can be arbitrary, in the
frame of the Hilbert transform (2).

The possibility for $\Gamma$ to be not small plays an important role
for the increasing of the range of validity of the kinetic Fermi
liquid equation. The comparison of the results obtained on the basis
of this equation with experimental data confirmed the validity of
the equation far beyond the limits established by the conditions
(36)--(38) and, consequently, for a essentially larger temperature
interval.

The $Ansatz$ 3 (16) leads to the elimination of the second Poisson
bracket in Eq. (11) in a very natural way. As it turned out, this
term in the generalized KB kinetic equation (11) is the main
obstacle on the way of the extension of Fermi liquid equation (40)
to a larger temperature interval \cite{23}. We should mention that
the validity of the kinetic equation (40) in the case of finite
values of $\Gamma$ was shown by different mathematical method in
\cite{30} but the result looked contradicting to the commonly
accepted conditions (36)-(38). Only the direct \lq\lq lawful"
elimination of the term $[\text{Re}\ g, \sigma^<]$ in Eq. (11) makes
the situation clear.

Despite the absence of the rigorous foundation of the validity of
the Fermi-liquid equations in the case of large $\Gamma$, these
equations were successfully used for the variety of systems of
strongly interacting particles. Thus, such equations were used for
theoretical description of the magneto-ordered state in 3d metals in
the framework of the Anderson periodic model \cite{31}. The value of
$\Gamma$ was considered to be constant and of the same order as
other energy parameters of the system: the width of the d level was
considered to be equal to a relatively large jump parameter $v : v
\sim\Gamma\sim1\ eV$. The obtained results turned to be in a good
agreement with  experimented data for the photoeffect on polarized
electrons \cite{32,33}. In conclusion, we will mention that the
$Ansatz$ 3 (16) can be preferable also for calculations of the
equilibrium properties such as produced for a nuclear matter in
\cite{12}. As it is stated in this work, using the extended
quasiparticle approximation (14), the total binding energy is
obtained to be
$$
E_{EQP}=17.4 \ \text{MeV/nucleon}.
$$
In contrast, the quasiparticle approximation (5) leads to the
result
$$
E_{QP}=15.9 \ \text{MeV/nucleon}.
$$
Bruckner theory meanwhile gives
$$E_B=16.7 \ \text{MeV/nucleon}.
$$
Due to the extra factor $Z^2(\vec{p}){[Z(\vec{p})<1]}$ in the second
term of the $Ansatz$ 3 compared with the $Ansatz$ 1, it is clear
that the result of this $Ansatz$ must be between the values of
$E_{QP}$ and $E_{EQP}$. It will be closer to the value of $E_B$ than
$E_{EQP}$ and closer to the value of $E_B$ than $E_{QP}$.

\end{document}